\documentclass{article}
\usepackage{spconf,amsmath,graphicx}
\usepackage{bm}
\usepackage{algorithm}
\usepackage{algorithmic}
\usepackage{array}
\usepackage{multirow}

\DeclareMathOperator*{\argmaxx}{argmax}

\title{ADDRESSING AMBIGUITY OF EMOTION LABELS THROUGH META-LEARNING}
\name{Takuya Fujioka, Dario Bertero, Takeshi Homma, Kenji Nagamatsu}
\address{Research \& Development Group, Hitachi, Ltd., Tokyo, Japan.}
\begin{document}

\ninept
%\tenpt
%
\maketitle
\begin{abstract}
Emotion labels in emotion recognition corpora are highly noisy and ambiguous, due to the annotators' subjective perception of emotions. Such ambiguity may introduce errors in automatic classification and affect the overall performance. We therefore propose a dynamic label correction and sample contribution weight estimation model. Our model is based on a standard BLSTM model with attention with two extra parameters. The first learns a new corrected label distribution, and is aimed to fix the inaccurate labels from the dataset. The other instead estimates the contribution of each sample to the training process, and is aimed to ignore the ambiguous and noisy samples while giving higher weight to the clear ones. We train our model through an alternating optimization method, where in the first epoch we update the neural network parameters, and in the second we keep them fixed to update the label correction and sample importance parameters. When training and evaluating our model on the IEMOCAP dataset, we obtained a weighted accuracy (WA) and unweighted accuracy (UA) of respectively 65.9\% and 61.4\%. This yielded an absolute improvement of 2.5\%, 2.7\% respectively compared to a BLSTM with attention baseline, trained on the corpus gold labels.
\end{abstract}
\section{Introduction}
\label{sec:intro}

Automatic recognition of affect and emotion is important to enable a more natural and engaging communication between humans and machines. In this work we concentrate on emotion recognition from speech, which is the task to estimate the emotional content of a spoken utterance.

In the past, emotion recognition was performed by extracting a set of low-level features from each frame of an audio sample. These features were then aggregated through various statistical aggregation function (mean, standard deviation, min, max, etc.) to a global utterance-level vector representation~\cite{schuller2010interspeech}, to be finally fed through a shallow classifier such as Support Vector Machines (SVM)~\cite{Schuller_2006, Alvarez_2006}. However, in recent years, the accuracy of speech emotion recognition has dramatically improved with the introduction of Deep Neural Networks (DNN). Initial DNN-based models~\cite{Stuhlsatz_2011} were still based on the same utterance-level feature extraction. However, in subsequent approaches, speech features extracted from each frame were used as inputs of more complex neural network architectures such as Convolutional Neural Networks (CNN) and Recurrent Neural Networks (RNN), and the accuracy was further improved~\cite{Han_2014, Lee_2015, Mirsamadi_2017}. Recent years saw the application of novel methods developed from other AI fields, such as self-attention models~\cite{Li_2019__}, Connectionist Temporal Classification (CTC)~\cite{Zhao_2019} and Dilated Residual Network (DRN)~\cite{Li_2019}. Even higher performance was achieved by employing multi-modal information, such as audio and image together with speech~\cite{Li_2019_}.

While most of the effort concentrated on the development of more accurate classification models, there were other aspects of emotion classification regarding the data itself that were mostly ignored, but that could help improving the performance. In many datasets, the emotional labels are annotated based on human annotators' perception and sensibility to emotion. Emotion perception is highly subjective~\cite{Douglas_2005}, therefore the labels often contain some noise due to humans' decision ambiguity. For instance, an annotator may assign the label \textit{neutral} not when the sample is actually neutral, but when he is unsure about the most correct emotion class. Likewise, he may mistakenly recognize some loud enthusiastic speech as angry, while instead it is happy. Training a model on such noisy labels is likely the cause of some performance degradation, because the model may become confused and may not clearly distinguish one emotion from another.

Another important issue is that, in many emotion recognition datasets, the numbers of utterances for each emotional category are imbalanced. Generally, in the classification task using these category imbalanced dataset, accuracy of the small class is decreased~\cite{Liu_2009, Sun_2012}, which in turn affects the overall accuracy. To overcome these problems, some methods were proposed to employ soft target approaches to correct the annotation ambiguities~\cite{Ando_2018}, or to augment the dataset with synthetic data to reduce the effect of data imbalance~\cite{Bao_2019}. However, the former method only performs a static label contribution estimation based on the original annotation data, while the latter method is complex and the generated data might still be affected by the original labeling noise. In other domains, such as image recognition, similar issues were tackled by performing the label update, not a priori but during training, by gradually tuning the estimation~\cite{Tanaka_2018}.

Inspired by the achievements in image recognition~\cite{Tanaka_2018}, we propose a method to automatically tune the contribution of each data sample during training.  We do this by alternately updating the parameters of a DNN emotion classification model, and then use the neural network prediction to correct the relative contribution and the target labels of each sample, in order to reduce the overall loss. The main purpose is to correct or ignore altogether the ambiguously labeled utterances, while giving higher importance to the clear and unambiguous ones. Results obtained in the Interactive Emotional Dyadic Motion Capture (IEMOCAP) dataset~\cite{Busso_2008} show that our proposed method is effective in removing the annotation noise. It achieves an improvement of 2.5\% for weighted accuracy, and of 2.7\% for unweighted accuracy compared to a state-of-the-art BLSTM model trained on the original labels only~\cite{Mirsamadi_2017}.

%We summarize our major contributions as follows:
%\begin{itemize}
%\item Introducing category labels updated to the emotion recognition task in order to eliminate ambiguous labels.
%\item Defining the contribution weights to the model learning for each utterance which are updated with emotional category labels.
%\end{itemize}

\section{Methodology}
\label{sec:method}

% \subsection{Summary}
Given an input audio sample ${\bf x}_{n}=[{\bf x}_{n, 1}, {\bf x}_{n, 2}, \cdot\cdot\cdot, {\bf x}_{n, T}]$, where $n$ is the utterance index, and ${\bf x}_{n, t}$ a frame-based feature vector, and $T$ the total number of frames, we want to estimate the probabilities of each emotion category ${\bf y}_{n}=[y_{n, 1}, y_{n, 2}, \cdot\cdot\cdot, y_{n, C}]$, where $C$ is the number of discrete emotion classes.
%from the feature sequence using DNN model with parameters $\bm{\theta}$;
%\begin{equation}
%{\bf y}_{n, c}=p(s_{n, c}|{\bf x}_{n}, \bm{\theta}),
%\end{equation}
We use a BLSTM model with attention~\cite{Mirsamadi_2017} to perform the classification.

To improve the classification performance, and reduce the ambiguity of the human-annotated labels during training, for each training speech sample we also learn two parameters: $\bm{L}_n=[l_{n, 1}, l_{n, 2}, \cdot\cdot\cdot, l_{n, C}]$, and $\bm{w}_n$. $\bm{L}_n$ represents a new estimate of each sample emotion class, aiming to correct the ambiguities and inaccuracies during the training process, while through $\bm{w}_n$ we learn a contribution weight for each training utterance. 

\subsection{Emotion classification model}
\label{sec:model}

\begin{figure}[t]
  \centering
  \centerline{\includegraphics[width=6.5cm]{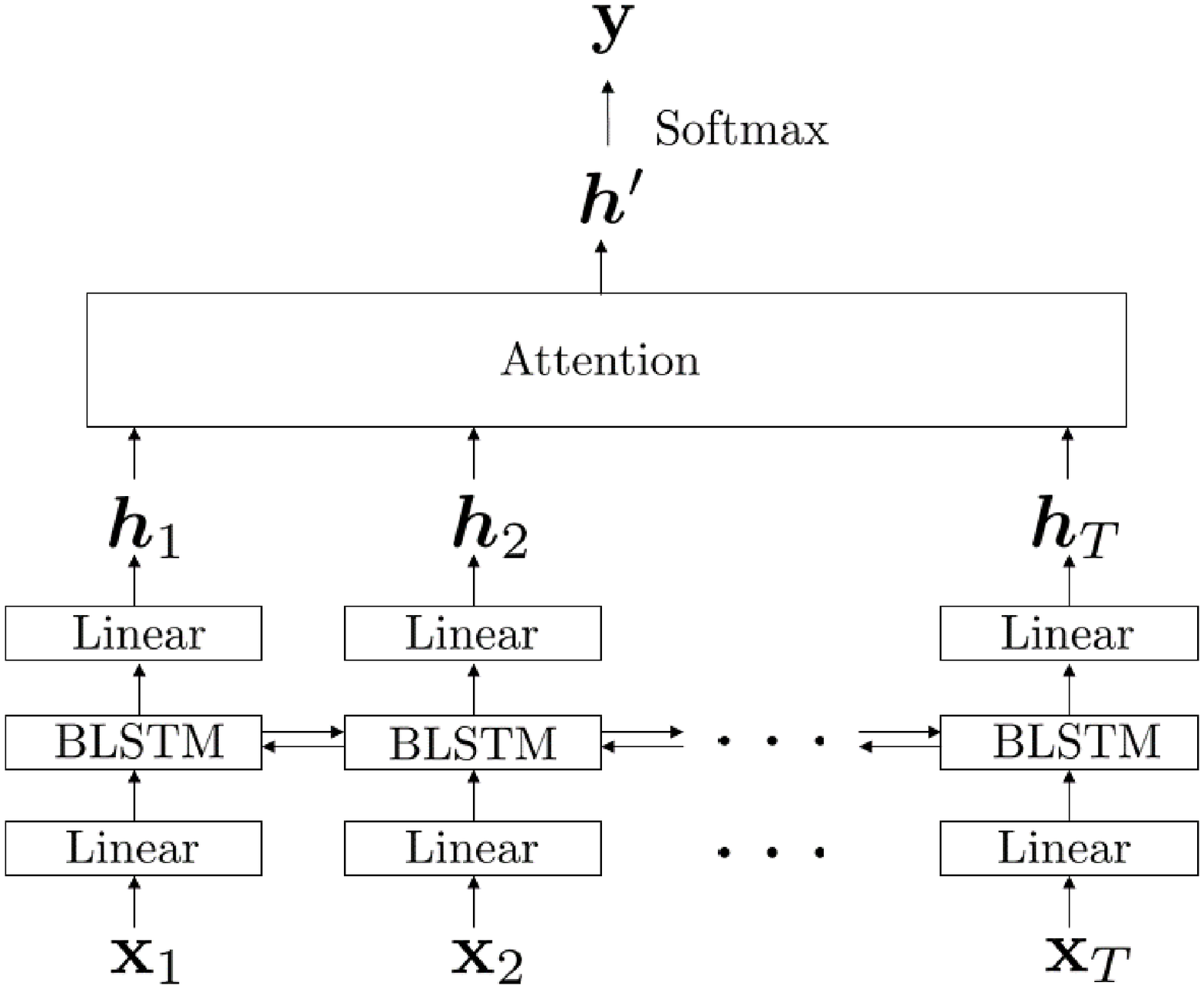}}
%  \vspace{-0.2cm}
\caption{The structure of the BLSTM model with attention.}\medskip
\label{fig:Mirsamadi}
%\vspace{-10pt}
%\vspace{-0.4cm}
\end{figure}

Fig.~\ref{fig:Mirsamadi} shows the structure of our main BLSTM emotion recognition model, which closely follows the state-of-the-art by \cite{Mirsamadi_2017}. At first, we input the feature sequence ${\bf x}_{n}$ through a Bi-directional LSTM (BLSTM), which yields $\bm{h}_{n}=[\bm{h}_{n, 1}, \bm{h}_{n, 2}, \cdot\cdot\cdot, \bm{h}_{n, T}]$ as the output. We then weight the contribution of each frame through an attention layer, where its weights $\alpha_{n, t}$ are calculated as follow:
\begin{equation}
\alpha_{n, t}=\frac{\exp(\bm{h}_{n, t}\bm{u}^{\top})}{\sum_{\tau=1}^{T}\exp(\bm{h}_{n, \tau}\bm{u}^{\top})}.
\end{equation}
In the equation above, $\bm{u}=[u_{1}, u_{2}, \cdot\cdot\cdot, u_{C}]$ are the learned attention parameters. The obtained attention weights $\alpha_{n, t}$ are used to calculate a weighted average over time of the BLSTM output vectors, in order to get a fixed-length utterance-level vector representation $\bm{h}^{\prime}_{n}$. We get the output emotion probabilities ${\bf y}_{n}$ by applying a softmax layer to $\bm{h}^{\prime}$:
\begin{equation}
\bm{h}^{\prime}_{n}=\sum_{t=1}^{T}\alpha_{n, t}\bm{h}_{n, t},
\end{equation}
\begin{equation}
y_{n, c}=\frac{\exp(h^{\prime}_{n, c})}{\sum_{c=1}^{C}\exp(h^{\prime}_{n, c})}.
\end{equation}

\subsection{Update of target labels and contribution weights}

\begin{figure}[t]
  \centering
  \centerline{\includegraphics[width=6.5cm]{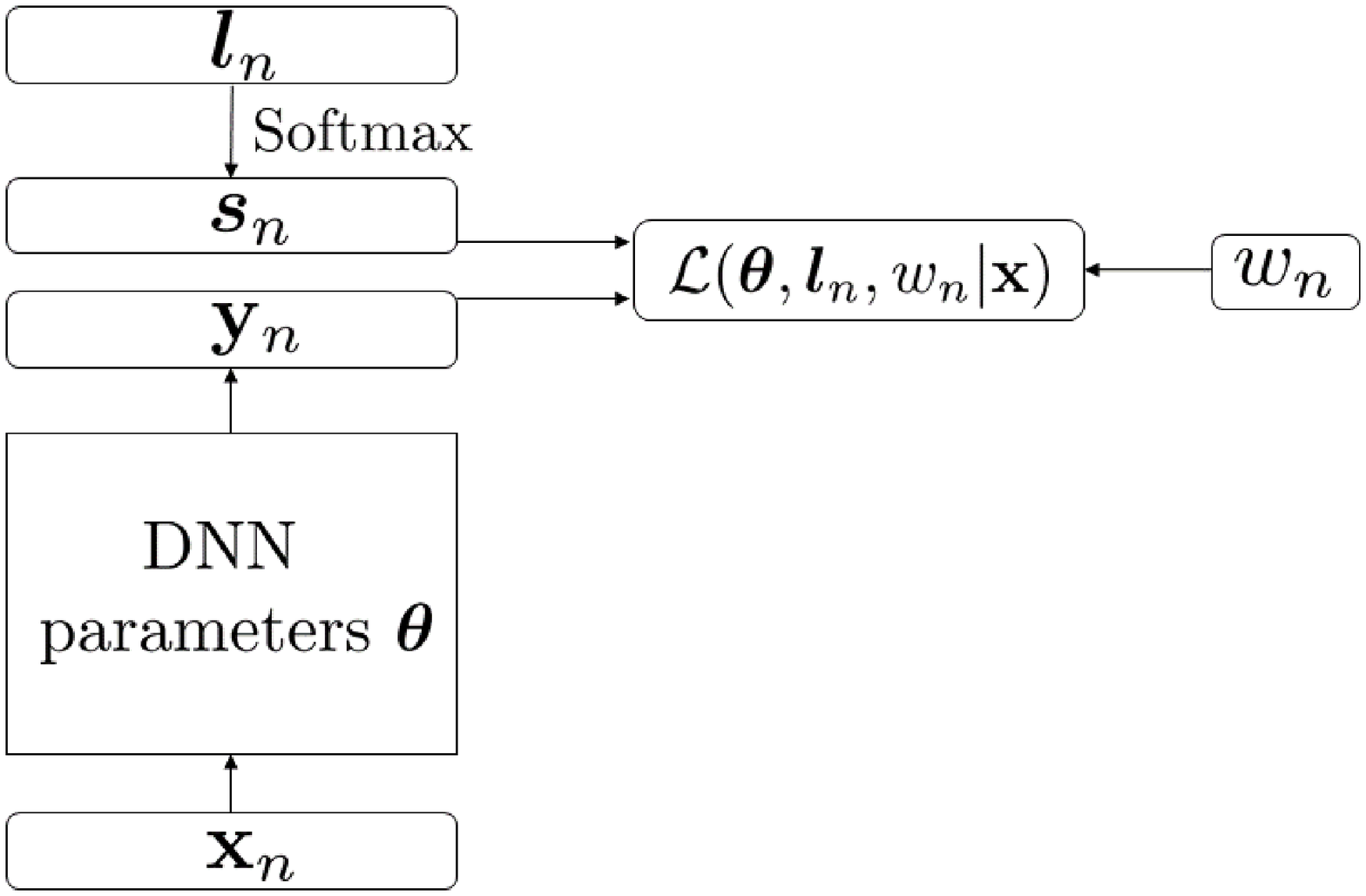}}
%  \vspace{-0.2cm}
\caption{Overview of the proposed training framework.}\medskip
\label{fig:learning}
%\vspace{-10pt}
%\vspace{-0.4cm}
\end{figure}

Normally, an emotion classification system such as the one explained in the previous section, is trained from the gold standard labels $\mathbf{y}_n$, with all the training samples having the same contribution weight during training. This is however not ideal for emotion recognition since the human annotated labels can be ambiguous or not precise. For instance, a sample can have been marked as ``neutral'' just because the annotators were unsure about the most appropriate emotion label, and not because it was actually neutral.

We therefore correct these inaccuracies and ambiguities by learning two extra parameters. The first one is $\bm{l}_{n}=[l_{n, 1}, l_{n, 2}, \cdot\cdot\cdot, l_{n, C}]\in\{0, 1\}$, inspired by~\cite{Tanaka_2018}. We initialize it with the one-hot gold standard emotion label. Through the learning process, this parameter is supposed to learn, for each training sample, the correct emotion distribution, eventually overriding the one previously assigned by the annotators. The second parameter, $w_{n}$, is instead a per-sample contribution weight. We initialize $w_{n}$ by applying the same method proposed in~\cite{Mirsamadi_2017}, by taking the proportion of samples in each emotion category:
\begin{equation}
w_{n}=\frac{\sum_{\nu=1}^{N}l_{\nu, c^{*}}}{\sum_{n=1}^{N}\sum_{\nu=1}^{N}l_{\nu, c^{*}}/N},
\end{equation}
\begin{equation}
c^{*}=\argmaxx_{c\in\{1, \cdot\cdot\cdot, C\}}\bm{l}_{n}.
\end{equation}
In~\cite{Mirsamadi_2017}, $w_{n}$ was proposed to address the class imbalance and prevent the recall degradation due to it, and was kept fixed throughout the training. We instead update it during training, as we assume that the model would learn to give higher weights to clear samples, and lower weights to ambiguous samples that presumably would only add noise to the classifier.

To update those parameters, and apply them in the classification process, we designed the framework shown in Fig.~\ref{fig:learning}. An overall model loss function is defined as $\mathcal{L}(\bm{\theta}, \bm{l}_{n}, w_{n}|{\bf x}_{n})$:
\begin{equation}
\mathcal{L}(\bm{\theta}, \bm{l}_{n}, w_{n}|{\bf x}_{n})=-\frac{\sum_{c=1}^{C}s_{n, c}\log{y_{n, c}}}{w_{n}},
\label{eq:loss}
\end{equation}
 where $\bm{s}_{n}$ is the mapping of $\bm{l}_{n}$ by softmax function to make it a probability distribution over emotions:
\begin{equation}
s_{n, c}=\frac{\exp{l_{n, c}}}{\sum_{\gamma=1}^{C}\exp{l_{n, \gamma}}}.
\end{equation}
It is worth noticing that in the cross entropy function we do not use the gold standard labels, but only the new learned emotion representation $\bm{l}_{n}$. 

The model parameters $\bm{\theta}$, $\bm{l}_{n}$ and $w_{n}$ are updated through an alternating optimization process, shown in Algorithm 1. Alternately, we first update $\bm{\theta}$, keeping  $\bm{l}_{n}$ and $w_{n}$ fixed, through one epoch. In the second step, we update $\bm{l}_{n}$ and $w_{n}$, keeping the BLSTM network weights fixed, through another one epoch. To avoid the algorithm to converge on very high values of $w_{n}$ when minimizing the loss function, we scale the value of $w_{n}$ after each update to maintain the following constraint:
\begin{equation}
\frac{\sum_{n=1}^{N}w_{n}}{N}=1.
\end{equation} 

\begin{algorithm}                      
\caption{Alternating optimization algorithm}         
\label{alg}                          
\begin{algorithmic}
\FOR{ $i\leftarrow 1$ to $num\_epochs$}
\STATE update $\bm{\theta}^{(i+1)}$ using $\bm{L}^{(i)}$ and $\bm{w}^{(i)}$
\STATE update $\bm{L}^{(i+1)}$ and $\bm{w}^{(i+1)}$ using $\bm{\theta}^{(i+1)}$
\ENDFOR
\end{algorithmic}
\end{algorithm}
%\vspace{-10pt}

\section{Experiments}
\label{sec:experiment}
\subsection{Corpus}
To evaluate the performance of the proposed learning method, we use the IEMOCAP dataset~\cite{Busso_2008}, one of the most commonly used benchmark datasets in emotion recognition tasks. The corpus is organized in 5 sessions, in each of which two actors performed a conversation. The total number of speakers in the corpus is 10. We only considered the samples belonging to the four emotional categories of \textit{happiness}, \textit{sadness}, \textit{neutral} and \textit{anger}, to keep the analysis consistent with previous works~\cite{Lee_2015, Mirsamadi_2017, Li_2019__, Zhao_2019, Li_2019, Li_2019_, Ando_2018, Bao_2019}. The number of utterances in each emotional class of each speaker is shown in Table~\ref{table:ut_num}. We performed a leave-one-speaker-out 10-fold cross-validation using a leave-one-out strategy~\cite{Li_2019}. We applied early-stopping criteria in all conditions to minimize the loss of the validation set~\cite{Ando_2018}.
\subsection{Experimental setup}

\begin{table}[tb]% Table 1
\caption{Number of the utterances in each emotional class and speaker.}
\label{table:ut_num}
\begin{center}
\small
\begin{tabular}{crrrr}
\hline
Speaker & \textit{Happiness} & \textit{Sadness} & \textit{Neutral} & \textit{Anger} \\ \hline
Ses01F & 69 & 78 & 171 &147 \\
Ses01M & 66 & 116 & 213 & 82 \\
Ses02F & 70 & 113 & 135 & 67 \\
Ses02M & 47 & 84 & 227 & 70 \\
Ses03F & 80 & 172 & 130 & 92 \\
Ses03M & 55 & 133 & 190 & 148 \\
Ses04F & 31 & 62 & 76 & 205 \\
Ses04M & 34 & 81 & 182 & 122 \\
Ses05F & 77 & 132 & 221 & 78 \\
Ses05M & 66 & 113 & 163 & 92 \\ \hline
Total & 595 & 1084 & 1708 & 1103 \\ \hline
\vspace{-20pt}
\end{tabular}%
\end{center}
\end{table}

\begin{table*}[tb]% Table 1
\caption{Results, percentage, over each method. P: Precision, R: Recall, F1: F1-score, WA: Weighted Accuracy, UA: Unweighted accuracy. *: reported values in the original papers.}
\label{table:accuracy}
%\vspace{-0.3cm}
\begin{center}
\small
%\footnotesize
\begin{tabular}{l >{\centering}p{8pt} >{\centering}p{8pt} >{\centering}p{8pt} c >{\centering}p{8pt} >{\centering}p{8pt} >{\centering}p{8pt} p{0.5pt} >{\centering}p{8pt} >{\centering}p{8pt} >{\centering}p{8pt} p{0.5pt} >{\centering}p{8pt} >{\centering}p{8pt} >{\centering}p{8pt}ccc}
\hline
 & \multicolumn{3}{c}{\textit{Happiness}} && \multicolumn{3}{c}{\textit{Sadness}} && \multicolumn{3}{c}{\textit{Neutral}}& & \multicolumn{3}{c}{\textit{Anger}} & &\\ 
\cline{2-4}\cline{6-8}\cline{10-12}\cline{14-16}
Method&P&R&F1&&P&R&F1&&P&R&F1&&P&R&F1&&WA&UA\\ \hline

BLSTM + ATT~\cite{Mirsamadi_2017} &38.8& 35.3 &35.2&& \textbf{64.5}& 68.2 &65.0&&\textbf{65.1}& 58.3 &60.9&&73.0& 76.1 &74.0&& 63.6 & 59.5 \\

BLSTM + ATT + Oversampling &42.6& 36.0 & \textbf{36.5} && 62.8 & 62.2 & 61.4 && 63.7 & 59.6 & 60.1 && 74.3 & 78.6 & 75.5 && 63.5 & 59.1 \\

BLSTM + ATT + Undersampling &35.4& 34.7 & 34.4 && 64.5 & 63.1 & 62.9 && 64.9 & 61.2 & 62.4 && 73.3 & 74.4 & 72.7 && 63.2 & 58.4 \\

Soft-target~\cite{Ando_2018} &N/A& N/A &N/A&& N/A& N/A &N/A&&N/A& N/A &N/A&&N/A& N/A &N/A&& 62.6* & \textbf{63.7}*\\

Cycle-GAN~\cite{Bao_2019} &N/A& \textbf{54}* &N/A&& N/A& 69* &N/A&&N/A& 51* &N/A&&N/A& 69* &N/A&& N/A & 60.4*\\

  \hline

BLSTM + ATT + $\bm{L}$ &44.9& 24.1 &28.4&& 62.8& 70.2 &65.3&&63.0& \textbf{66.4} &\textbf{63.9}&&75.9& 75.9 &75.1&& 64.7 & 59.2\\

BLSTM + ATT + $\bm{w}$ &45.4& 30.5 &34.7&& 61.9&71.7&65.3&&63.8& 63.7 &63.1&&\textbf{78.3}& 77.2 &\textbf{77.2}&& 65.2 & 60.8\\

BLSTM + ATT +  $\bm{L}$ + $\bm{w}$ pretrained &51.0& 22.6 &29.4&& 62.7& 68.8 &65.1&&58.9& 65.4 &61.6&&75.6& 74.9 &74.7&& 64.4 & 57.9\\
  
BLSTM + ATT +  $\bm{L}$ + $\bm{w}$ &\textbf{53.4}& 30.2 &35.7&& 62.8& \textbf{74.1} &\textbf{67.1}&&64.1& 63.7 &63.3&&75.7& 77.6 &75.9&& \textbf{65.9} & 61.4\\ \hline

\vspace{-15pt}
\end{tabular}%
\end{center}
\end{table*}

\begin{table*}[tb]% Table 1
\caption{Mean values of the contribution weights $w_{n}$ before and after training in the \textbf{BLSTM + $\bm{L}$ + $\bm{w}$}. A lower value means higher importance to the loss function. The model assigns a much lower importance to the \textit{happiness} label, which is presumably the most ambiguous also given the lower overall performance.}
\label{table:weight}
%\vspace{-0.3cm}
\begin{center}
\small
\begin{tabular}{ccccc}
\hline
 & \multicolumn{4}{c}{Mean values of the contribution weights $w_{n}$} \\
\cline{2-5}
& \textit{Happiness} & \textit{Sadness} & \textit{Neutral} & \textit{Anger} \\ \hline
%Label change ratio [\%] & 42.7 & 12.1 & 21.1 & 17.0 \\\hline
Initial $w_{n}$& 0.53 & 0.97 & 1.52 & 0.98 \\ %\cline{2-6}
%Trained $w_{n}$ for initial emotional labels& 1.36 & 0.85 & 1.05 & 0.88 \\ \hline
%Initial $w_{n}$ for trained emotional labels& 0.68 & 1.02 & 1.39 & 1.02 \\ %\cline{2-6}
Learned $w_{n}$& 1.32 & 0.92 & 1.08 & 0.84 \\ \hline
\vspace{-20pt}
\end{tabular}%
\end{center}
\end{table*}

We extracted 32-dimensional acoustic features from the raw audio samples using openSMILE toolkit~\cite{Eyben_2010}; 12-dimensional Mel-Frequency Cepstral Coefficients (MFCCs), loudness, fundamental frequency ($F_{0}$), voicing probability, zero-crossing rate, and the first order derivatives of them. The frame length and frame shift were set to 25 ms and 10 ms, respectively. All features were normalized by mean and standard deviation calculated over all of the utterance features in the training set.

The emotion classification model was composed of a fully-connected layer with Rectified Linear Unit (ReLU), a BLSTM layer and a fully-connected layer. The numbers of hidden units were 512, 128 and 4, respectively. We applied dropout to all the layers, with a dropout rate was 0.5. We used Adam~\cite{kingma2014adam} as an optimization algorithm.

We evaluated our model using two common evaluation measures in the previous works: weighted accuracy (WA) and unweighted accuracy (UA). We also calculated per-class precision, recall and F1-score, in order to get a performance estimate over each individual emotion class.

We compare our model (\textbf{BLSTM + $\bm{L}$ + $\bm{w}$}) against the following baselines:
\begin{itemize}
\item \textbf{BLSTM + ATT}: our reimplementation of the attention based BLSTM model as proposed in~\cite{Mirsamadi_2017}.
\item \textbf{BLSTM + ATT + Oversampling/Undersampling}: same as the above, but applying oversampling or undersampling instead of $\bm{w}$ to address the class imbalance problem.
\item \textbf{BLSTM + ATT + $\bm{L}$}: the full model and training algorithm, but only updating $\bm{L}$, while keeping $\bm{w}$ fixed.
\item \textbf{BLSTM + ATT + $\bm{w}$}: the full model and training algorithm, but only updating $\bm{w}$, while keeping $\bm{L}$ equal to the gold standard labels.
\item \textbf{BLSTM + ATT +  $\bm{L}$ + $\bm{w}$ pretrained}: similar to the full model, but the neural network was first pretrained until hitting the early stopping condition on the validation set. %The motivation of this learning condition is same as~\cite{Tanaka_2018}.
\item \textbf{Soft-target}: the soft label method proposed in~\cite{Ando_2018}.
  \item \textbf{Cycle-GAN}: the data augmentation method proposed in~\cite{Bao_2019}.

\end{itemize}

\subsection{Results}

\begin{figure}[t]
  \centering
  \centerline{\includegraphics[width=6.5cm]{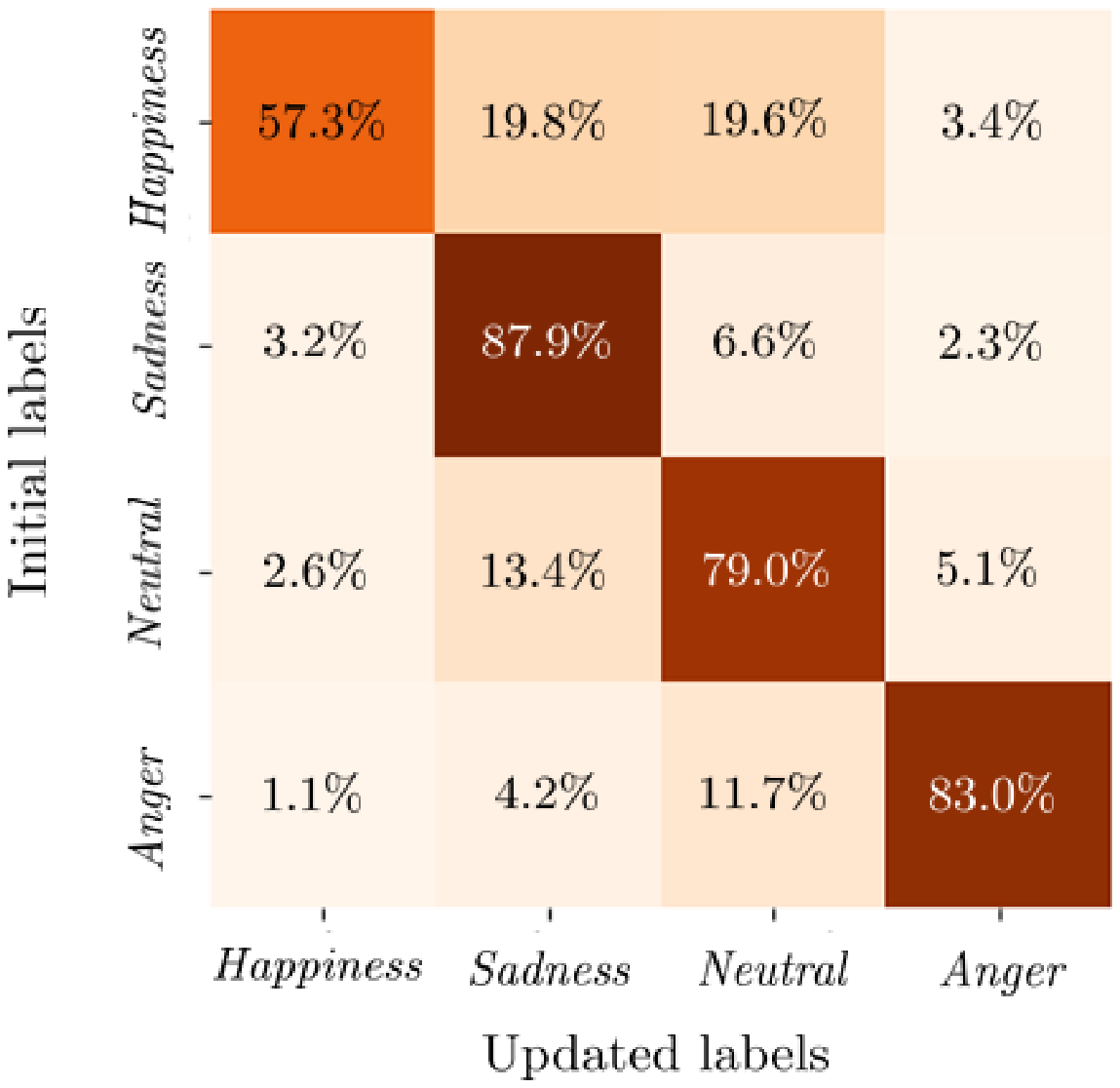}}
%  \vspace{-0.2cm}
\caption{Percentages of labels update from and to each emotional category in the \textbf{BLSTM + $\bm{L}$ + $\bm{w}$}. Most of the updates affect the \emph{happiness} class, which is often changed to \emph{sadness} or \emph{neutral}.}\medskip
\label{fig:heatmap}
%\vspace{-10pt}
%\vspace{-0.4cm}
\end{figure}

Final results are shown in Table~\ref{table:accuracy}. For the Soft-target and Cycle-GAN baseline we show the reported results from the original papers, since in the former case we were unable to replicate the same results, and the latter used a different methodology. Our proposed model performed the best in terms of WA, and the second best in terms of UA, achieving 65.9\% and 61.4\%, respectively. This yields an absolute improvement of respectively +2.5\% and +2.7\% over the BLSTM + ATT baseline. It is also worth noticing that the original reported WA and UA in~\cite{Mirsamadi_2017} are 63.5\% and 58.8\%, respectively. These values are not significantly different from the ones obtained by our reimplementation. The soft-target baseline achieved an higher UA than our proposed method by +2.3\%. However, they only used one-fold cross validation instead of ten-fold, and they did not report the performance for the individual emotion classes~\cite{Ando_2018}, therefore the results are not fully comparable.

In terms of per-class performance, our model achieves an F1-score of 35.7\%, 67.1\%, 63.3\% and 75.9\% respectively for the \textit{happiness}, \textit{sadness}, \textit{neutral} and \textit{anger} classes, with absolute improvements of +0.5\%, +2.1\%, +2.4\% and +1.9\%, respectively. The lower improvement in F1-score in \textit{happiness} is compensated by a significant improvement in precision of +14.6\%.

\subsection{Discussion}
By looking at the results in Table~\ref{table:accuracy}, it clearly emerges how our proposed model achieves a much better performance than just applying some simple imbalance corrections such as data undersampling or oversampling. In terms of performance, the introduction of the per-sample importance weighting $\bm{w}$ had a slightly higher influence than the emotion correction parameter $\bm{L}$, presumably because it is less sensible to errors. $\bm{w}$ had the main effect of improving the precision on \textit{happiness}, and of improving the precision in \textit{anger}, while the main contribution of $\bm{L}$ was to improve the recall on \textit{neutral} samples. Pretraining the model with the original labels did not seem to work better than starting immediately updating $\bm{L}$ and $\bm{w}$, presumably due to a greater learning bias over incorrect and ambiguous gold labels.

It is interesting to notice how these two parameters affect the various emotion classes after training. Table~\ref{table:weight} shows the change of $\bm{w}$, a lower value means a greater weight of the loss function in eq.~\ref{eq:loss}. The weight given to \textit{happiness} samples, initially the less numerous class, was greatly reduced during training. By looking at the final precision and recall on this class, this seems a consequence of the very high ambiguity of the \textit{happiness} annotations, that classifiers have a great difficulty in distinguishing and clearly separate from other classes.
% The issues observed in the \emph{happiness} class are consistent with what was reported in the previous work~\cite{Bao_2019}.

Likewise, we observed a similar behavior regarding the $\bm{L}$ parameter. Figure~\ref{fig:heatmap} shows the amount of label updates as learned by $\bm{L}$ during training. Only in around half of the cases the label \textit{happiness} was kept, while it was often changed into \textit{sadness} or \textit{neutral}. Besides \textit{happiness}, in around 10\% of the cases, \textit{anger} was updated to \textit{neutral}, while \textit{neutral} was updated to \textit{sadness}. These latter changes are likely due to the aforementioned subjectivity of the emotion, and of the boundaries between them, which are leading to ambiguous choices.

\section{Conclusion}
\label{sec:conclusions}

We have proposed a novel meta-learning approach, built on top of a traditional BLSTM with attention classifier, to address the issue of labeling inaccuracy and ambiguity in speech emotion recognition. Our proposed method is effective for dynamically update each sample label during training, and learn an estimate of each sample contribution to reduce the relative weight of ambiguous utterances. We obtained an overall performance of 65.9\% and 61.4\%, respectively weighted and unweighted accuracy, on the IEMOCAP dataset, giving an absolute improvement of 2.5\% and 2.7\%, respectively over the same BLSTM model trained on the original gold labels. We also showed how our proposed framework clearly managed to reduce the importance of the most ambiguous label (\textit{happiness}), and to fix the initial label annotation to the most appropriate classes for each sample, thus improving the classification performance.

%\vfill
%\pagebreak
\clearpage

\bibliographystyle{IEEEbib}
\bibliography{strings,refs}

\end{document}